\documentclass[aps,prl,twocolumn,superscriptaddress,showpacs,floatfix]{revtex4}
\usepackage{graphicx}
\usepackage{color}
\usepackage{bm}
\usepackage{amsmath}

\newcommand{\eq}[1]{Eq.~(\ref{#1})}
\newcommand{\fig}[1]{Fig.~\ref{#1}}
\newcommand{\be}[1]{\begin{equation}\label{#1}}
\newcommand{\ee}{\end{equation}}

\newcommand{\aga}[1]{{\color{red}\it #1}}
\renewcommand{\vec}[1]{{\boldsymbol #1}}

\begin{document}

\title{Initial state dependence in multi-electron threshold ionization of 
atoms}

\author{Agapi Emmanouilidou}
\affiliation{ITS, University of Oregon, Eugene, Oregon 97403-5203} \date{\today}

\author{Peijie Wang\footnote{present address: The Beijing Key Laboratory for 
Nano-Photonics and Nano-Structure,
Capital Normal University,
Beijing 100037 P.R. China }}
\affiliation{Max Planck Institute for the Physics of Complex Systems, N{\"o}thnitzer
Str.\ 38, D-01187 Dresden, Germany}
\author{Jan M.\ Rost}
\affiliation{Max Planck Institute for the Physics of Complex Systems, N{\"o}thnitzer
Str.\ 38, D-01187 Dresden, Germany}

\date{\today}

\begin{abstract}
	It is shown that the geometry of multi-electron threshold ionization
	in atoms depends on the initial configuration of bound
	electrons.  The reason for this behavior is found in the
	stability properties of the classical fixed point of the
	equations of motion for multiple threshold fragmentation.
	Specifically for three-electron break-up, apart from the
	symmetric triangular configuration also a break-up of lower
	symmetry in the form of a T-shape can occur, as we demonstrate
	by calculating triple photoionization for the lithium ground
	and first excited states.  We predict the electron break-up
	geometry for threshold fragmentation experiments.
\end{abstract}

 \maketitle 

Three-body Coulomb dynamics, in particular two-electron atoms, are
very well studied in the energy regime of single as well as double
ionization \cite{bych+07,brsc00,mrbi+99,tari+00} .  Much less is known about
correlated dynamics in four body Coulomb systems, more precisely on
differential observables for fragmentation of a three-electron atom
in its nucleus and all electrons \cite{copi06,gowa+06,em07}.  A recent
experiment provides for the first time detailed information in terms
of differential cross sections on the angular and energetic break-up
parameters of three electrons following impact double ionization of
Helium \cite{dudo+07}.  For small excess energies $E$ (each continuum
electron carries away about 9 eV energy), it was found that the
electrons form an equilateral triangle upon breaking away from the
nucleus.  This is expected in accordance with Wannier's theory
\cite{wa53}, quantified for three electrons in \cite{klsc76}.  There,
it is shown that the fixed point of classical dynamics, through which
full fragmentation near threshold $E=0$ should proceed, is given for a
three-electron atom by an equilateral triangle with the nucleus in the
center and the electrons at the corners.

In two-electron atoms the corresponding fixed point implies a
collinear escape of the electrons in opposite directions
\cite{wa53,ro94}.  The normal mode vibration about this collinear
configuration is stable, which is in marked contrast to the
three-electron case, where the triangular configuration is linked to
two unstable, degenerate normal modes \cite{klsc76,kuos98}.

We will show that as a consequence of this property the preferred 
final geometry of the three escaping electrons becomes initial state 
dependent and can change between an equilateral triangle 
and  a less symmetric T-shaped escape. While the former is realized, 
e.g., in electron impact double ionization of helium,  the latter should be 
seen in triple photoionization of lithium.  These are only two 
prominent examples, the general pattern and the reason for it will be 
detailed below. 

Due to the scaling of the Coulomb potential, states of finite total
angular momentum $L$ will all behave like the $L=0$ state close to
threshold which is therefore sufficient to consider \cite{ro98}.  In
hyperspherical coordinates with the radial variable $w$ instead of 
the hyperradius $r = w^{2}$, the Hamiltonian for a three-electron atom
with total angular momentum $L=0$ reads \be{ham} h =
\frac{p_{w}^{2}}{8w^{2}}+\frac{\Lambda^{2}}{2w^{4}}+\frac{C(\vec\Omega)}{w^{2}}
\ee
where 
$\vec\Omega=(\alpha_{1},\alpha_{2},\theta_{1},\theta_{2})^{\dagger}$
contains all angular variables describing the
positions of the electrons on the hypersphere of radius $r$
and $\Lambda$, the so called grand angular momentum operator
\cite{sm60}, is a function of $\vec\Omega $ and all conjugate momenta.  The
total Coulomb interaction $V =C/r$ acquires in this form simply an
angular dependent charge $C(\vec\Omega)$.  In terms of the familiar
vectors  $\vec r_{1},\vec r_{2},\vec r_{3}$,
pointing from the nucleus to each electron, the hyperspherical
coordinates are given by
\begin{subequations}
    \label{hyper}
\begin{eqnarray}
    r &=&
(r_{1}^{2}+r_{2}^{2}+r_{3}^{2})^{1/2}\\
\alpha_{1} &=& \arctan(r_{1}/r_{2})\\
\alpha_{2}&=&\arctan (r_{2e}/r_{3})\\
\theta_{1}&=&\arctan[\vec r_{1}\cdot \vec
r_{2}/(r_{1}r_{2})]\\
\theta_{2}&=&\arctan[\vec r_{1}\cdot\vec r_{3}/(r_{1}r_{3})]\,,
\end{eqnarray}
\end{subequations}
where $r_{2e}= (r_{1}^{2}+r_{2}^{2})^{1/2}$.

Threshold dynamics is governed by motion along the normal modes about
the fixed point of the Hamiltonian $H =(h-E)w^{2}$ \cite{ro98a}.  The
special form of $H$ ensures that the dynamics remains regular
approaching the fixed point radially, i.e., $w\to w^{*}=0$, while
$\vec\Omega^{*}$ at the fixed point is defined through
$\nabla_{\Omega}C(\vec\Omega)|_{\vec\Omega=\vec\Omega^{*}}=0$.  
 The 
equations of motion can be expressed as a system of
first order differential equations (ODE) $\dot{\vec\Gamma} = {\cal 
G}\nabla_{\vec\Gamma}H$ for the phase space vector 
$\vec\Gamma = (p_{w},\vec P_{\vec\Omega},w,\vec\Omega)^{\dagger}$, 
with
\be{gmatrix} {\cal G}=\left (\begin{array}{cc} 0
& -{\mathbf 1}_{f}\\
{\mathbf 1}_{f} & 0 \end{array}\right )\, 
\ee 
 a block matrix
composed from 0 and unity matrices of dimension $f\times f$
\cite{al88in}, where $f$ is the number of degrees of freedom, here
$f=5$.  Since the differential equations are still singular at the
fixed point $(w^*,\Omega^{*})$, a change of the momentum variables $\vec P_{\Omega}$
conjugate to $\Omega$ is needed
\be{momenta} p_{\omega_j}= P_{\omega_j}/w \ee as well as a new time
variable $\tau$ related to the original time $t$ conjugate to the Hamiltonian
\eq{ham} through $dt = w^{3}d\tau$.  Finally, the normal modes can be
obtained from the modified ODE, $d\vec\gamma/d\tau ={\cal
G}\nabla_{\gamma}\tilde H$ by diagonalizing the matrix
$\partial^{2}{\cal G}\tilde
H/(\partial\vec\gamma\partial\vec\gamma)|_{\vec\gamma=\vec\gamma^{*}}$,
where $\vec\gamma$ refers to the new phase space variables with the
(non-canonical) momenta from \eq{momenta}. The eigenvalues are the
Liapunov exponents $\lambda_{j}$ and in the normal mode basis $\{\hat
u_{j}\}$ threshold dynamics assumes an oscillator-like form of
$\delta\vec u_{j}(\tau)=\exp(\lambda_{j}\tau)\delta\vec u_{j}(0)$
with the unit vectors $\hat u_{j}$
\be{unitvector} 
\hat u_{j} = \delta\vec u_{j}(0) /| \delta\vec u_{j}(0)|
\ee
defining the normal mode basis.
 The $\delta\vec\gamma(\tau)=\vec\gamma -\vec\gamma^{*}$ are excursions of
$\vec \gamma$ from their fixed point values $\vec \gamma^{*}$ and are expressed 
as a linear combination of the $\delta\vec u_{j}(\tau)$.

We recall briefly the familiar three-body break-up in a two-electron
atom with hyperradius $r_{2e}$ and the angles $\alpha_{1}, \theta_{1}$
defined as in \eq{hyper}.  The charge corresponding to $C(\Omega)$ in
\eq{hyper} is for the two-electron problem
\be{c2e} C_{2e}(\alpha_{1},\theta_{1})=
-\frac{Z}{\sin\alpha_{1}}-\frac{Z}{\cos\alpha_{1}}
+\frac{1}{(1-\sin(2\alpha_{1})\cos\theta_{1})^{\frac12}}\,.  \ee The
fixed point analysis reveals a pair of unstable
$\lambda_{1/2}=-\lambda_{0}\pm\lambda$ and stable
$\lambda_{3/4}=-\omega_{0}\pm i\omega$ Liapunov exponents
$(\lambda_{0},\omega_{0},\lambda,\omega>0)$ with a shift
($-\lambda_{0}$ and $-\omega_{0}$, respectively), compared to standard
symplectic dynamics.  The shift formally arises through the
noncanonical transformation of the momentum variables \eq{momenta}
necessary to obtain normal mode motion about the singular fixed point
$w^{*}=0$.

The resulting eigenvectors $\hat u_{i}$ reveal orthogonal motion along
$\theta_{1}$ and $\alpha_{1}$, i.e., any phase space vector 
$\delta\vec\gamma_{\alpha_{1}}(\tau)$
describing linearized motion in the subspace spanned by 
$\vec p_{\alpha_{1}},\vec\alpha_{1}$, can be expressed as a linear 
combination of two eigenvectors
 $\delta\vec\gamma_{\alpha_{1}} = a_{\gamma_{\alpha}}\exp(\lambda_{1}\tau)\hat
u_{1}+b_{\gamma_{\alpha}}\exp(\lambda_{2}\tau)\hat u_{2}$.
An analogous relation holds for linearized motion in the subspace
$\vec p_{\theta_{1}},\vec\theta_{1}$, realized through two {\it 
different} eigenvectors,
$\delta\vec\gamma_{\theta_{1}} =
a_{\gamma_{\theta}}\exp(\lambda_{3}\tau)\hat u_{3}
+b_{\gamma_{\theta}}\exp(\lambda_{4}\tau)\hat u_{4}$. 
Hence, 
$\delta\vec\gamma_{\theta_{1}}(\tau)\cdot\delta\vec\gamma_{\alpha_{1}}(\tau)=0$ at
all times $\tau$.
The coincidence of the eigenspaces with the respective dynamics of
$\theta_{1}$ and $\alpha_{1}$ has the important consequence that the
fixed point value $\theta_{1}=\pi$ is preserved through its relation
to the stable eigenmode while all energy sharings can occur through
the relation of the unstable eigenmode with $\alpha_{1}$.

This coincidence of normal modes with subspaces of observables is
special to two-electron atoms and does not hold for more electrons.
Moreover, for three-electron atoms, a new feature emerges in threshold
dynamics, namely the existence of two degenerate pairs of unstable normal modes
with Liapunov exponents $\lambda_{1/2}=\lambda_{3/4}=-\lambda_{0}\pm\lambda$.  
Note that the fragmentation dynamics 
close to the fixed point will take place
in the phase space of the two unstable normal modes with an arbitrary
linear combination of the two  eigenvectors $\hat u_{1}$
and $\hat u_{3}$ belonging to the two (equal) positive Liapunov 
exponents $\lambda_{1}=\lambda_{3}=\lambda_{+}$,
\be{fragmentation}
\delta\vec\gamma(\tau)=\exp(\lambda_{+}\tau)(c_{1} \hat u_{1}+c_{3} \hat 
u_{3})\,,
\ee
which allows for flexibility in the four-body
break-up, as we will see.

\begin{table}
    \caption{\label{table1} 
    The two eigenvectors $\hat u_{i}$ belonging to 
    the positive Liapunov exponent $\lambda>0$ in the basis of the 
    phase space variables from \eq{hyper}}.\\ 
        \begin{tabular}{l||c|c}
	\hline
	 {basis}& $\hat u_{1}$ & $\hat u_{3}$\\\hline
	  $\delta \alpha_{1}$& $-8.22\times 10^{-3}$ &$ 2.67\times 10^{-1}$\\
	  $\delta  p_{\alpha_{1}}$&$ -2.97\times 10^{-2}$ &$ 9.64\times 10^{-1}$\\
	  $\delta \alpha_{2}$&$ -1.81\times 10^{-1}$ & 0\\
	  $\delta  p_{\alpha_{2}}$&$ -9.83\times 10^{-1}$ & 0\\
	  $\delta \theta_{1}$&$ 7.12\times 10^{-3}$ & 0\\
	  $\delta  p_{\theta_{1}}$&$6.57\times 10^{-3}$ &$ -4.46\times 10^{-3}$\\
	  $\delta  \theta_{2}$&$3.33\times 10^{-3}$ &$ 7.40\times 10^{-3}$\\
	  $\delta  p_{\theta_{2}}$&$-2.75\times 10^{-4}$ &$ 8.92\times 10^{-3}$
	  \\\hline\hline
    \end{tabular}
\end{table}

One sees directly from Table I that normal mode dynamics about the
fixed point for geometrical angles $\theta_{i}$ and hyperangles
$\alpha_{i}$ is not separated as for the two electron case, and even
more importantly, all phase space variables are linked to the unstable
normal modes.  Hence, the fixed point geometry does not provide
necessarily a preference for the final angles of the electrons.  On the
other hand, this opens the way for the initial state to have an
influence on the final observables, even close to threshold.

Relevant for threshold ionization is the spatial electron distribution
at the time (we label it $\tau = 0$) when all electrons to be ionized
have received enough energy (through collisions) to leave the atom.
We call this distribution the transient threshold configuration (TTC).

In a two electron-atom, the necessary energy transfer between the two 
electrons leading to three-body fragmentation happens through a 
single collision. At this time $\tau=0$ both electrons are naturally 
close together, so that $\delta\alpha_{1}(0) \approx 0$ holds and defines the 
TTC, independent of the initial bound electron  configuration.

In a three-electron atom, the situation is more complicated since at
least two collisions are necessary to distribute the energy among the
electrons so that all of them can escape.  For triple photo-ionization
of Lithium we know from classical calculations that the 1s photo
electron (3), which has absorbed the photon initially, collides
immediately with the other 1s electron (2) and subsequently (about 60
attoseconds later) either electron 2 or the photo electron itself
collides with the 2s electron (1).  This can be expressed with the two
collision sequences $s_{1}=(32, 21)$ and $s_{2}=(32, 31)$
\footnote{Note that for the present purpose we have relabeled the
electrons compared to \cite{emro06}.}.  The time delay of the second
collision, respectively, is due to the ``distance'' of the 2s shell
from the 1s shell.  It leaves an asymmetric situation after the second
collision when the transient threshold configuration is reached at
$\tau = 0$.  While the two electrons participating in the last
collision are close to each other, the third one is further away.
Concentrating on the collision sequence $(32, 21)$, this implies in
terms of distances to the nucleus \be{TTC} r_{1}\approx r_{2}\ne
r_{3}\,.  \ee Hence, $\delta\alpha_{1}(0)\approx 0$ while
$\delta\alpha_{2}(0)\ne 0$.
This can be easily accommodated with a suitable linear combination 
choosing, e.g.,  $c_{1} =1/a$ and $c_{2} =-1/b$ in
\eq{fragmentation}, where $a$ and $b$ are the coefficients of 
$\delta\alpha_{1}$ with respect to $\hat u_{1}$ and $\hat u_{3}$  
(see first row of Table I).

The scenario described is still within the overall picture of
threshold break-up in the spirit of Wannier and therefore in
accordance with the power law $\sigma \propto (E/E_{0})^{\alpha}$ for
the dependence of the total break-up cross section on the excess
energy $E$ \cite{saan88,wepa+00,bllu+04}.  The exponent $\alpha =
(\lambda_{1}+\lambda_{3})/(2\lambda_{r}) = 2.162$, where
$\lambda_{r}=2.5088$ is the radial Liapunov exponent and
$\lambda_{1}=\lambda_{3}=5.4240$.  Yet, the TTC with unequal distances
of the electrons to the nucleus breaks the complete symmetry among the
electrons.  This in turn, leads to a preferred final geometry of the
three electrons since essentially with the third electron being far
away, the geometry determination reduces to that of two electrons
escaping in an equivalent way as in the two-electron break-up, i.e.,
back-to-back with an angle of $\theta_{1}=\pi$.  The question remains
if there is a preferred angle between this electron pair escaping
along a line and the third electron.  Assuming for simplicity that
$r_{3}\gg r_{1}\approx r_{2}$ and therefore
$\sin\alpha_{2}\approx 0$ holds (the opposite case would
lead to the same result), we can expand $C(\Omega)$ in \eq{ham} in
powers of $\eta\equiv\alpha_{2}$
\be{pot23}
C(\Omega) \approx \eta^{-1}\sum_{n=0}^{3}c_{n}\eta^{n}
\ee
with
\begin{subequations}
    \label{pot23a}
\begin{eqnarray} 
c_{0}&=&C_{2e}(\theta_{1},\alpha_{1})\\
c_{1}&=&2-Z\\
c_{2}&=& 
\cos\theta_{2}\sin\alpha_{1}+\cos\theta_{12}\cos\alpha_{1}\\
c_{3}&=&{\textstyle\frac{3}{2}}(\cos\theta_{2}\sin\alpha_{1})^{2}+
{\textstyle\frac{3}{2}}(\cos\theta_{12}\cos\alpha_{1})^{2}\nonumber\\
&&+(1-2Z)/6
\,,
\end{eqnarray}
\end{subequations}
where $\theta_{12}=\theta_{2}-\theta_{1}$.  To lowest order in $\eta$,
the problem to find a stable configuration is that of the two-electron
system (here with nuclear charge $Z=3$) with the well known solution
$\theta_{1}^{**}=\pi$ and $\alpha_{1}^{**}=\pi/4$ \cite{wa53}.  These
values minimize $c_{2}$ for any value $\theta_{2}$.  Its value
$\theta_{2}^{**}=\pi/2$ is determined from $\partial
c_{3}/\partial\theta_{2}=0$ which is a stable solution.  This
constitutes the T-shape of the three electrons as a preferred
asymptotically ($\eta\to 0$) stable geometrical configuration within
the globally unstable two-dimensional subspace spanned by $\hat
u_{1},\hat u_{3}$. The fragmentation dynamics in the sub space is 
completely degenerate - hence, we have looked on a ``higher order'' 
correction, which would give a preference within the degenerate 
subspace: that is the asymptotically stable T-shape. 
 Indeed, this configuration was found numerically
close to threshold in triple photo ionization \cite{emro06}.

One can double check this new insight into the role of the TTC of the
electrons in the presence of degenerate Liapunov exponents by
considering initial configurations which lead to different transient
configurations.

For an initial excited state Li($1s2s^{2}$) the two collisions in
which energy is transferred from the (1s) photo electron to the two 2s
electrons happen close in time (and therefore in space).  Hence, the
transient threshold configuration after the second collision at $\tau
= 0$ is $r_{1}\approx r_{2}\approx r_{3}$ close to the fixed point and
therefore giving preference to the symmetric break-up with a dominant
angle 120$^{\circ}$ between the electrons, similar to electron impact
double ionization of Helium.
 
\begin{figure}[tb]
      \includegraphics[width=0.99\columnwidth]{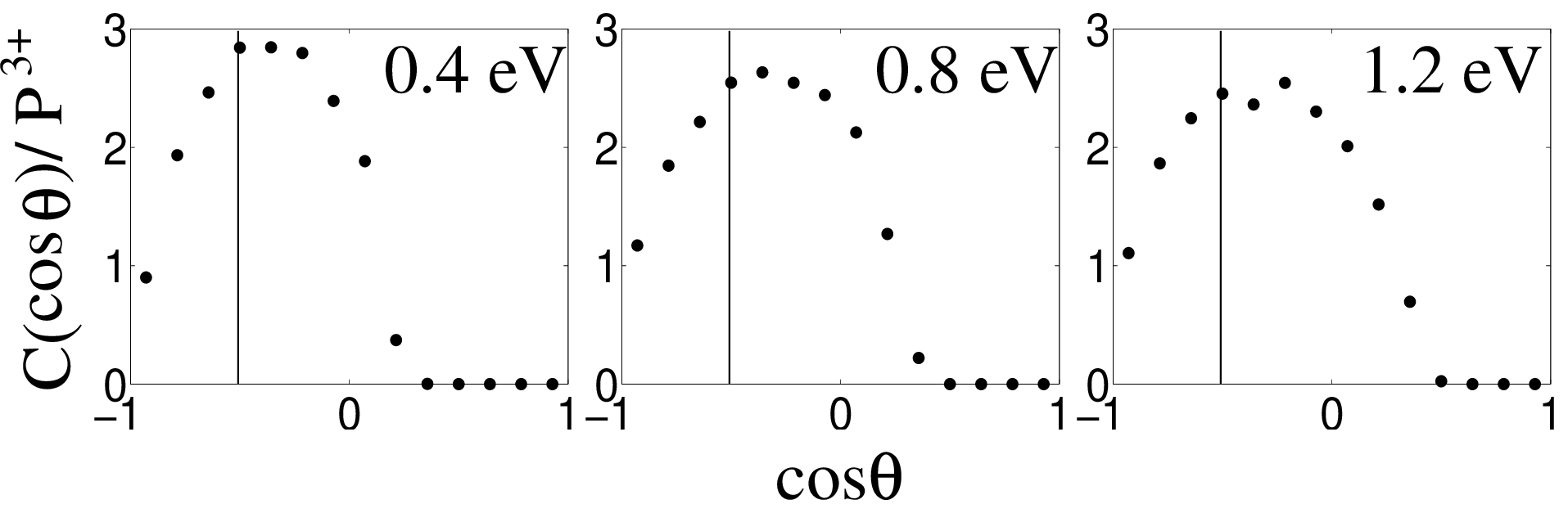}
      \caption{Probability to find the angle $\theta$ between two 
      electrons for triple ionization from the initial Li($1s2s^{2}$) 
	   state for excess energies above the triple ionization threshold 
	   as indicated.}
	   \label{excited}
       \end{figure}

Numerical results  confirm this prediction as seen from \fig{excited}
where the probability to find an angle $\theta$ between two electrons 
in triple photo ionization of  the initial Li($1s2s^{2}$) is shown 
for different excess energies. Clearly, the most likely angle is 
$\cos\theta = -1/2$ corresponding to $120^{\circ}$, indicated by the 
thin vertical line.

\begin{figure}[tb]
	\includegraphics[width=0.99\columnwidth]{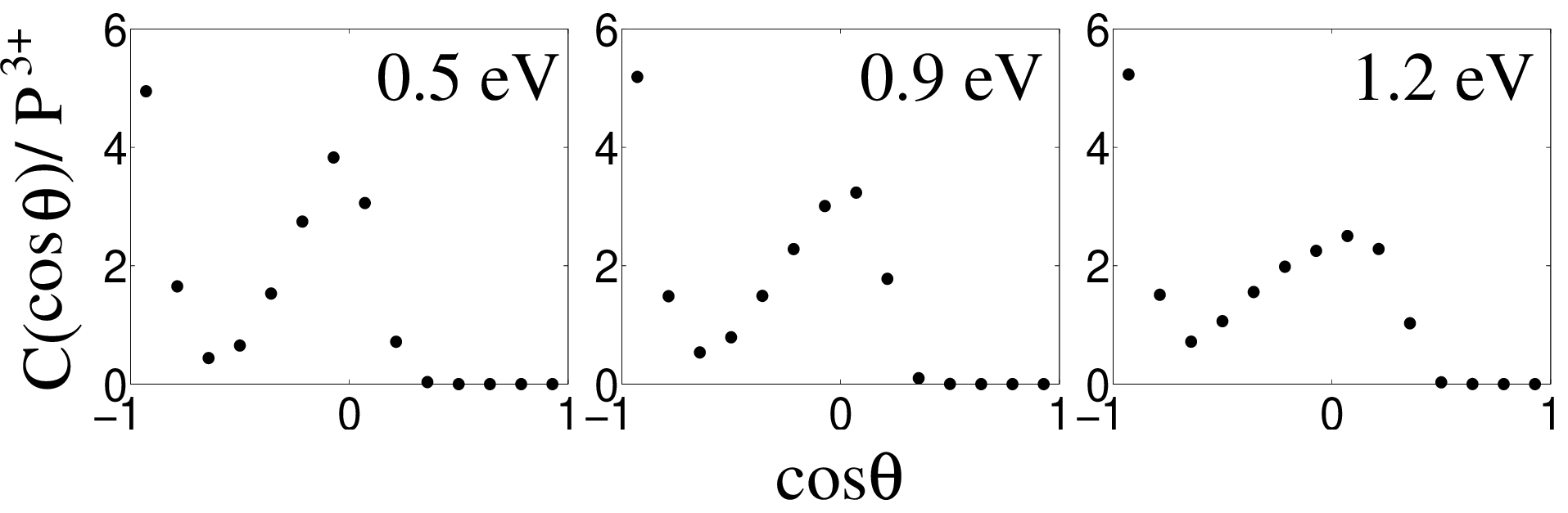}
	\caption{Same as \fig{excited}, but for triple ionization 
	from the initial Li($1s^{2}2s$) 
	ground state for excess energies above the triple ionization threshold 
	as indicated.}
	\label{ground}
    \end{figure}

\begin{figure}[tb]
	   \includegraphics[width=0.44\columnwidth]{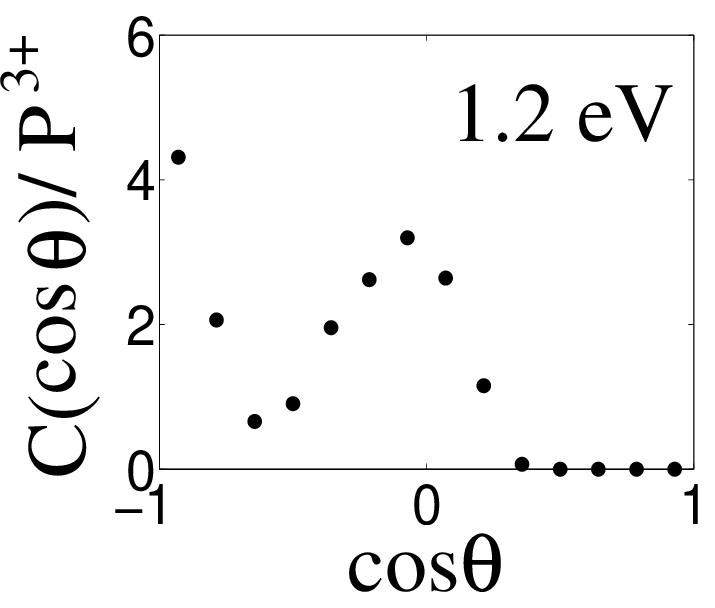}
	   \caption{Same as \fig{excited}, but with a 2s electron as 
	   photo electron.}
	   \label{excited-2}
       \end{figure}

On the other hand, for the ground state as initial state the TTC is
asymmetric as previously described and we expect a final T-shape
geometry with peaks at 90$^{\circ}$ and 180$^{\circ}$.  This is indeed
the case, as can be seen in \fig{ground}.  As a final test we may use
again the excited initial state Li($1s2s^{2}$) but take the 2s
electron as the photo electron.  This process is due to the smaller
dipole coupling by more than an order of magnitude suppressed compared
to the ionization with the 1s electron absorbing the photon.  However,
here we use this only as an illustration for the initial state
dependence of threshold ionization.  According to our reasoning we
have in this case (although it is the same initial state as before) a
different TTC since the first collision of the photo electron 1
happens with the 1s electron 2 while later on the collision of the 1s electron 2 with the
2s electron 3 will take place.  Consequently, for the TTC
$r_{3}\approx r_{2}\ne r_{1}$ holds which is structurally identical to
\eq{TTC}, the situation when a final T-shape geometry appears.  As can
be seen in \fig{excited-2} the T-shape indeed also emerges if 
 the photo electron comes from the 2s shell.

       \begin{table}
	   \caption{\label{table2}The relation of excitation process, initial 
	   configuration and geometry of the electrons for various 
	   three-electron break-up processes close to threshold.}
	   \begin{tabular}{|c|c|c|}
	       \hline
	       impact & initial  &  break-up\\
	       by & configuration &geometry\\\hline\hline
	       electron  & $1s^{2}$ (He) & $\triangle$\\
	       electron  & $1s2s$ (He) & $\perp$\\
	       photon  & $1s^{2}2s$ (Li) & $\perp$\\
	       photon  & $1s2s^{2}$ (Li) & $\triangle$\\
	       photon  & $1s2s3s$ (Li) & $\perp$\\
	       \hline\hline
	   \end{tabular}
       \end{table}

To summarize, we predict that for three electron break-up near
threshold two preferred geometrical patterns for the electrons exist,
namely an equilateral triangle and a T-shape.  Which of them is
realized depends on the transient threshold configuration, that is the
spatial distribution of the electrons at the time when the energy
among them is distributed such that all of them can escape.  The
transient threshold configuration is strongly influenced by the
initial state of the electrons, as has been discussed in detail for
the case of lithium.  A complete overview of the three-electron
break-up pattern for three electron systems is provided in Table
\ref{table2}.  Whenever the two electrons to which energy is
transferred in the course of the fragmentation of the atom are in the
same shell, a symmetric triangular geometry is expected.  If the two
electrons are from different shells, we expect a T-shape.  Recent
experimental results on electron impact double ionization of helium
\cite{dudo+07} are consistent with our predictions but do not provide
sufficient information to consider our prediction as experimentally
already confirmed.  Based on the present consideration it is also
possible to predict the break-up geometries and their realization
dependent on the initial state for more than three electrons.


\end{document}